\documentclass[12pt]{elsart}
\usepackage{epsfig}

\usepackage{amssymb}
\usepackage{amsfonts}
\usepackage{amsmath}

\newcommand{\bra}[1]{\ensuremath{\langle #1 |}}
\newcommand{\ket}[1]{\ensuremath{|\, #1 \rangle}}
\newcommand{\bk}[2]{\ensuremath{\langle #1 | #2 \rangle}}
\newcommand{\kb}[2]{\ensuremath{| #1 \rangle\!\langle #2 |}}

\begin{document}
\title{Entanglement of distant flux qubits mediated by non-classical
electromagnetic field}
\author{E Zipper$^1$, M Kurpas$^1$, J Dajka$^1$ and M Ku\'{s}$^2$ }
  
\address{$^1$ Institute of Physics, University of Silesia, Ul. Uniwersytecka 4, 40-007 Katowice, Poland}
\address{$^2$ Center for Theoretical Physics, Polish Academy
  of Sciences, Al. Lotnik\'{o}w 32/46, 02-668 Warszawa, Poland} 
\ead{kurpas@server.phys.us.edu.pl}

\begin{abstract}
The mechanism for entanglement of two flux qubits each interacting with a
single mode electromagnetic field is discussed. By performing a Bell state measurements (BSM) 
on photons we find the two qubits in an entangled state depending on the system parameters. 
 We discuss the results for two initial states and take into consideration the influence of decoherence.
\end{abstract}
\maketitle

\section{Introduction}
Entanglement is one of the most fundamental features of quantum mechanics.
Besides its fascinating conceptual aspect it also plays an important role in
quantum information science because entanglement of qubits is the essential
requirement for quantum computing. Various systems have been considered as
qubits \cite{raimond, nakamura}, among them the solid state ones seem to be
very promising. In particular superconducting flux qubit has been developed in
superconducting ring with Josephson Junction \cite{orlando,migliore}. The
junction playing the role of the tunneling barrier can be replaced by a
superconducting quantum wire which allows for quantum phase slip \cite{mooij}.
Recently a flux qubit based on semiconducting quantum ring with a controllable
barrier has been proposed \cite{zipper}. In this context the problem of
entanglement of two (or more) solid state qubits is of great importance. It has
been investigated for superconducting flux qubits interacting via the mutual
inductance, via the connecting loop with Josephson Junction and via the LC
circuit \cite{berkeley, plourde, paternostro, migliore}. It was found
\cite{berkeley} that entangled states do not decohere faster than the uncoupled
states. This is remarkable considering the expectation that spatially extended
entangled states could be very susceptible to decoherence.

\begin{figure}[ht]
\centering
\includegraphics [scale=0.3]{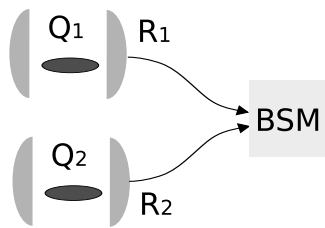}
\caption{Entanglement swapping scheme}
\label{fig1}
\end{figure}

In this paper we want to study the entanglement of distant flux qubits by
swapping. Presented model considerations may be applied both to superconducting
or semiconducting flux qubits. We investigate two independently evolving
subsystems each composed of a qubit exposed to a single mode of quantized
electromagnetic field (Fig. \ref{fig1}). Contrary to the previous studies where the so
called external approximation was used \cite{dajka} in this paper we
take into account the full qubit-field interaction.

The entanglement swapping was originally proposed for photons \cite{ekert} and
has been investigated both theoretically and experimentally \cite{riedmatten,
pan}. Recently this idea has been used to demonstrate the entanglement of two
single atom quantum bits each spontaneously emitting a photon \cite{monroe}. In
our paper we use this idea to entangle solid state qubits which seem to be the
most scalable and integrable \cite{kok}. The process of entanglement can be
described in this case by the interaction Hamiltonian with controllable
parameters. The use of solid state qubits instead of atomic qubits described in
\cite{monroe} allows to build systems operating at microwave rather than
optical frequencies.

The scheme of entanglement swapping for the discussed system is presented in Fig. \ref{fig1}.
Each qubit $Q$ interacts with an electromagnetic field mode $R$ leading to an
entangled qubit-field $(QR)$ state. This effect has been observed in a series
of experiments \cite{chio}. The two $(QR)_i$ ($i=1,2$) systems do not interact
with each other and therefore the state of the whole system is a product state.
If one then performs the Bell State Measurement (BSM) on $R_1$ and $R_2$, the
partner subsystems $Q_1$ and $Q_2$ will collapse to an entangled state although
they have never physically interacted. To enhance the qubit-field interaction
the qubits can be placed into the quantum cavity. The photons can escape from the
cavity e.g. through a less reflecting mirror \cite {kok, brune}. To quantify
the entanglement we calculate the negativity, we discuss the results
for two different initial states.\\
In chapters II-IV we investigate the behaviour at short time scales where the decoherence 
effects are negligible,  the influence of decoherence is studied in chapter V.
\section{The qubit-cavity system}
To show the idea we consider the rf-SQUID qubit \cite{migliore} in the presence
of static magnetic flux $\phi^{cl}$. The Hamiltonian $H_Q$ of such qubit can be
written in a pseudo-spin notation
\begin{equation}
H_Q=-\frac 12B_z{\sigma _z}-\frac 12B_x{\sigma _x}.
\label{HQ}
\end{equation}
We operate at $T \ll B_x/k_B$ in order to neglect thermal fluctuations.
The diagonal term $B_z$ in (\ref{HQ}) has the form
\begin {equation}
B_z = 2 I_c  \sqrt{6 (\beta_L -1)}(\frac{\phi_0}{2}-\phi)
\end{equation}
where $\beta_L=2 \pi L \left(I_c / \phi_0 \right) > 1$,$\phi = \phi^{cl}$, $I_c$ is the
Josephson junction critical current, $B_{x}$ is the tunneling energy between
the two potential wells. Close to $\phi =\frac 12\phi _0$ ($\phi_0=h/2e$) the ring is well described by the quantum superpositions of two
opposite
persistent current states. \\
At first we describe the process of entanglement of a qubit $Q_1$($Q_2)$ with a
single electromagnetic field mode $R_1(R_2)$. We model the electromagnetic
field of the resonant cavity as an LC resonator described by $H_{R}$
\begin{equation}
H_R=\hbar \omega_R\left( a^{\dagger }a+\frac 12\right).
\end{equation}
When the qubit is exposed to the quantized electromagnetic field the total flux
$\phi =\phi ^{cl}+\phi ^q,$ contains the quantum part
\begin{equation}
\phi^q = \sqrt{\frac{\hbar}{2 \omega_R C_R}} \left( a + a^{+}\right)
\end{equation}
 which leads to the qubit-field coupling.
After some algebra we obtain
\begin{eqnarray}
H_{QR} &= &\frac{\hbar \omega _{Q}}2\sigma _z+\hbar \omega _{R}\left(
a^{\dagger
}a+\frac 12\right)  \nonumber \\
&& -\hbar \tilde{g}\left( a+a^{\dagger } \right) \left(\sigma_z \cos\theta - \sigma_x
\sin \theta \right) \label{HQR}
\end{eqnarray}
where\\
$\omega _{Q}$ is the qubit frequency
\begin{equation}
 \frac{\hbar \omega _{Q}}2= \frac 12\sqrt{ \left( B_z^{cl}\right)^2
+B_x^2 },
\end{equation}
the "mixing angle" $\theta$ \cite{blais} is
\begin{equation}
\theta =\tan ^{-1}\frac{B_x}{B_z^{cl}},
\end{equation}
and the coupling constant $\tilde{g}$ takes the form
\begin{equation}
\tilde{g}=I_c \sqrt{\frac{3  \left(  \beta_L-1\right) }{\hbar\omega_R C_R}}.
\end{equation}

The above considerations can be equally well performed for a semiconducting
flux qubit \cite{zipper} with
\begin{equation}
B_{z} =2 I_0 \left( \frac{\phi_0}{2} - \phi^{cl} \right),
\end{equation}
where $I_0$ is the amplitude of persistent current, $\phi_0= h/e,$ $B_x$
describes the tunneling amplitude of an electron via a potential barrier.

 Assuming realistic values of the parameters for superconducting qubit  e.g.
$\omega _R=2\pi\cdot 50GHz$, $I_c=0.5\mu A$, we get $\tilde{g}=0.2\omega _R$.

To discuss the qubit-field entanglement we assume that the coherent coupling
overwhelms the dissipative processes (strong coupling regime). For creation and
manipulation of entangled states, it is thus essential that both the cavity
decoherence time $T_R$ and the qubit decoherence time $T_Q$ are much longer
than the qubit-cavity interaction time $T_\Omega \sim \pi / \tilde{g}
\sim10^{-11}$ s. Recently a high quality cavities (quality factor $Q_f\sim
10^5-10^8$) have been built \cite{brune, blais}. They have a photon storage
time $T_R$ in the range $0.3\mu s-300\mu s$. The estimated
decoherence times $T_Q$ of the considered qubits are of the order of a few $%
\mu s$ (to be specific we assume $T_Q\sim 1\mu s$ \cite{migliore}). In the next
two chapters  we investigate the system at $t\ll T_Q,T_R$ allowing the
entanglement to be obtained before the relaxation processes set in.

\section{Entanglement swapping}
The $(QR)_i$, $i=1,2$ system is described by a state vector $\left\vert
\psi_{QR}(t)\right\rangle_i $, which at $t=0$ is a direct product of the qubit
and the cavity states:
\begin{eqnarray}\label{rhopsi}
\rho_{(QR)_i}(0) &=&\vert \psi _{QR}(0)\rangle_{ii}\langle
\psi _{QR}(0)\vert,
\nonumber \\
\vert \psi _{QR}(0)\rangle_i &=&\vert \sigma n\rangle_
i =\vert \sigma \rangle_i \otimes \vert n\rangle_i,
\label{psi}
\end{eqnarray}
where $\sigma$ represents the qubit pseudo-spin states ($g$-ground ,$%
e$-excited), $\vert n\rangle$ are the photon number eigenstates, forming the so
called Fock basis, $%
n=0,1,2,...$.

The interaction of the qubit with the field
leads, in general, to the entangled state
\begin{equation}
\vert\psi_{QR}\left(t\right)\rangle_{i}= e^{-\frac{i}{\hbar}H_{QR}t} \vert\psi _{QR}\left(0\right)\rangle_i
\end{equation}
As the two qubit-boson subsystems do not interact with each other their time
evolved state remains separable:
\begin{equation}
\rho \left(t\right)=\rho_{(QR)_1}(t)\otimes \rho_{(QR)_2}(t)
\end{equation}
The time evolution of this composite is a product of two unitary evolutions of
its constituents generated by the Hamiltonian (\ref{HQR}) where
\begin{eqnarray}
\vert \psi _{QR}(t)\rangle_1 &=&\sum_n[a_n(t)\vert g n\rangle_1
+b_n(t)\vert e n\rangle_1 ] \\
\vert\psi _{QR}(t)\rangle_2 &=&\sum_n[\tilde{a}_n(t)\vert g n\rangle_2
+\tilde{%
b}_n(t)\vert e n\rangle_2]
\end{eqnarray}
The BSM is performed on electromagnetic modes in Fock basis (one photon with
the vacuum) \cite{riedmatten} and projects the formerly independent qubits onto
an entangled state

\begin{equation}\label{rhoqq}
\rho_{QQ}(t)=Tr_R\left(\vert B_R^1\rangle \langle B_R^1\vert \rho(t)\right) ,
\end{equation}
where
\begin{equation}\label{bell}
\vert B_R^1\rangle=\frac 1{\sqrt{2}}\left(\vert 01\rangle -\vert 10\rangle
\right)
\end{equation}
is one of the Bell states of the electromagnetic field modes, the trace $Tr_R$
is
taken with respect to photonic degrees of freedom. \\
After the BSM, the final qubit-qubit ($QQ$) state is of the form
\begin{equation}
\begin{array}{rcl}
  \vert \psi _{QQ}\rangle &=& [a_0(t)\tilde{a}_1(t)-a_1(t)\tilde{a}_0(t)]\vert gg\rangle \\
 &+&[a_0(t)\tilde{b}_1(t)-a_1(t)\tilde{b}_0(t)]\vert ge \rangle  \\ 
 &+&[b_0(t)\tilde{a}_1(t)-b_1(t)\tilde{a}_0(t)]\vert eg \rangle  \\ 
 &+&[b_0(t)\tilde{b}_1(t)-b_1(t)\tilde{b}_0(t)]\vert ee \rangle
\end{array}
\label{pqq}
\end{equation}

We quantify the entanglement by the {\it negativity} \cite{neg}
$N(\rho)=\max(0,-\sum_i \lambda_i)$, where $\lambda_i$ are negative
eigenvalues of the partially transposed \cite{peres} density matrix of the two
qubits. For an entangled state, the negativity is positive reaching its maximal
value $N=0.5$ for maximally entangled pure state. It vanishes for disentangled
states. Moreover, as it is an entanglement monotone it  can be used to quantify
the degree of entanglement. The use of negativity, instead of some entropic
criteria as e.g. linear entropy,  allows for simultaneous treatment of the
entanglement of pure and mixed states. Let us notice that in general (e.g.
beyond Jaynes--Cummings approximation)  the qubit--resonator system evolves in
an infinite dimensional Hilbert space. It is known \cite{hor4} that in  high
dimensional systems the so called PPT (positive with respect to partial transposition)
 entangled states can occur. They  cannot be
detected by the Peres criterion and negativity. In this paper we limit our attention
to the NPT entangled states i.e those which are negative with respect to partial transposition. 

\section{Numerical results}
We present results for entanglement of both qubit-field ($N_i$) and qubit-qubit
($N_{QQ}$) systems. As the calculations are numerical we are not limited to the
weak coupling regime. In numerical calculations the Hilbert space of microwave
modes is truncated at $n_{\max }=10$. We test the validity of the truncation by
controlling the traces of the matrices \cite{vourdas} being never smaller than
$0.99$.

There are many parameters affecting entanglement of qubits. To show the idea we
restrict our considerations to selected examples and discuss the results for
two initial states. In our model calculation we assume that both qubits are
identical, the analysis can easily be extended  beyond . In this paper we
consider only the resonant case i.e. $\omega_{R_i}=\omega_{Q_i}\equiv
\omega_R=2 \pi \cdot 50 GHz$. The values of $\tilde{g}_i$ are in the units of $\omega_{R}$.

\begin{figure}[ht]
\centering
\includegraphics[width=0.7\textwidth]{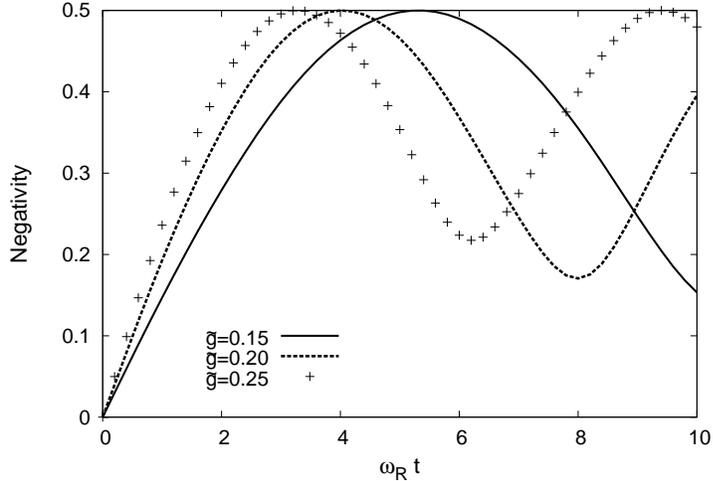}
\caption{The qubit-field negativity for different values of $\tilde{g}$,
$\theta=\pi/2$, initial state $\vert e0\rangle$ and $\omega_{Q}=\omega_R=2 \pi\cdot 50 GHz$.}
\label{fig2}
\end{figure}

\begin{figure}[ht]
\centering
\includegraphics[width=0.7\textwidth]{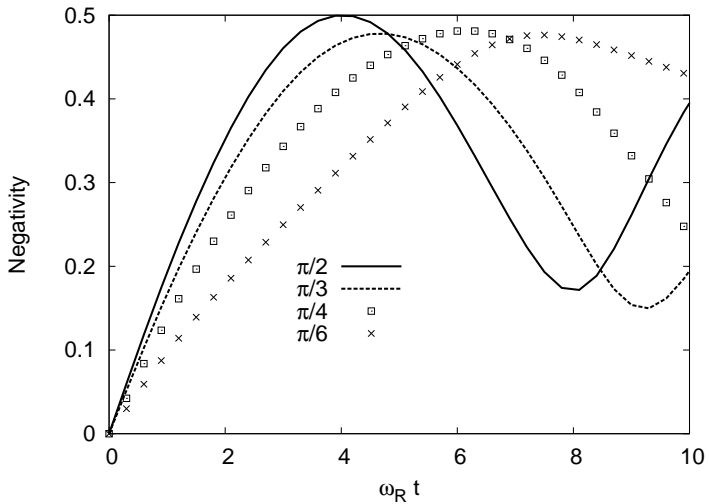}
\caption{The qubit-field negativity for different values of $\theta$,
$\tilde{g}=0.2$, initial state $\vert e0\rangle$, $\omega_{Q}=\omega_R=2 \pi\cdot 50 GHz$.}
\label{fig3}
\end{figure}

At first we assume the initial state to be
\begin{equation}
\vert\psi _{QR}(0)\rangle_1 \otimes\vert\psi _{QR}(0)\rangle_2=\vert e
0\rangle_1 \otimes\vert g 1\rangle_2. 
\label{e0g1}
\end{equation}
In Fig. \ref{fig2} we show how the qubit-field negativity depends on the coupling
strength $\tilde{g}$ and in Fig. \ref{fig3}  its behaviour for different values of the mixing
angle $\theta$.
Comparing these figures we see that both $\theta$ and $\tilde{g}$  influence the
effective qubit-field interaction strength. The increase of $\tilde{g}$ causes the
increase of the Rabi oscillation frequency and the entanglement arises faster
than for weaker coupling. Similarly, bringing $\theta$ closer to
$\pi/2$ increases the Rabi frequency. For $\theta=0$ the $QR$ entanglement disappears.
 In the following we assume $\theta=\pi/2$ which
gives the strongest effective coupling with fixed $\tilde{g}$.
\begin{figure}[ht]
\centering
\includegraphics [width=0.7\textwidth]{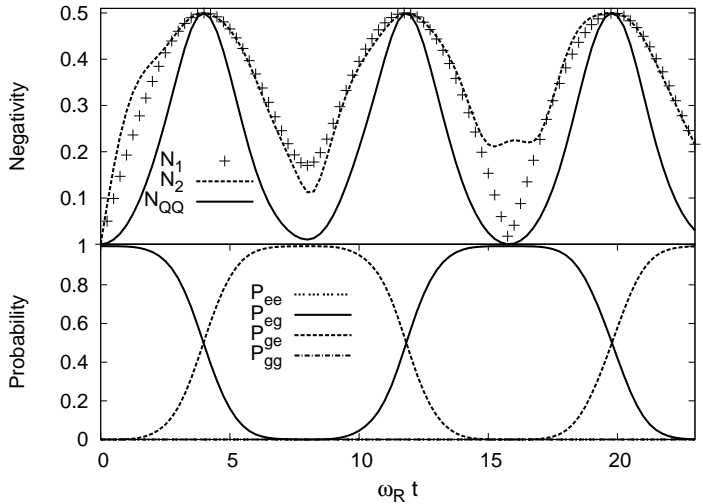}
\caption{ Negativities $N_1$ (crosses), $N_2$ (dashed line), $N_{QQ}$ (solid line) and probabilities for finding the two
qubits in different states after the BSM. The initial state $\vert e0\rangle_1 \otimes \vert g1\rangle_2$, coupling strength $\tilde{g}_i=0.2$, $\theta_i= \pi/2$.}
\label{fig4}
\end{figure}\\
In the upper panel of Fig. \ref{fig4} the oscillating qubit-field negativities $N_1$ and
$N_2$ reflect the varying degree of entanglement as a function of time. The
differences in these two curves arise from different initial states for
$(QR)_i$ systems ( $\vert e0\rangle_1 \otimes \vert g1\rangle_2$). If we perform the BSM at certain time $t$ we obtain an
entanglement of qubits (solid line) conditioned by the degree of entanglement
of ($QR)_i$. In particular if we do the BSM at the time window in which the
$(QR)_i$ subsystems are almost maximally entangled we obtain the  maximally
entangled qubits with $N_{QQ} \sim 0.5$. On the other hand if we perform the
BSM in the time window where the $(QR)_i$ subsystems are weakly entangled the
$QQ$ entanglement is vanishingly small.

We emphasize that the 'time' in the figures is either the physical time of the quantum
evolution of the $QR$ system or the time, called the 'BSM time', at which the BSM
was performed.

\begin{figure}[h]
\centering
\includegraphics [width=0.7\textwidth]{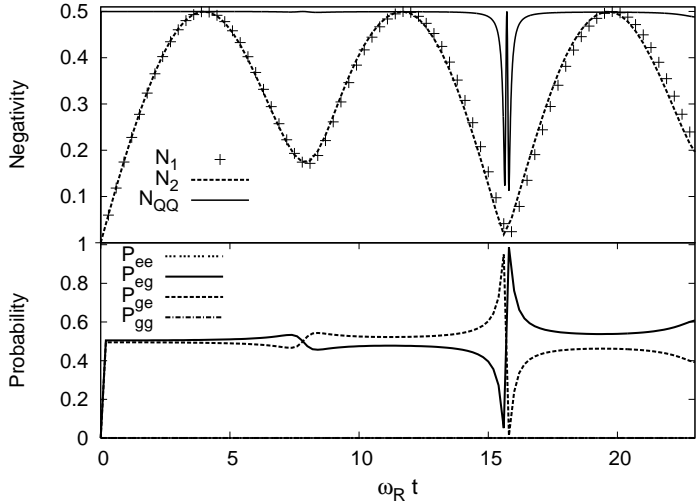}
\caption{$N_1$ (crosses), $N_2$ (dashed line) and $N_{QQ}$ (solid line) negativities (top) 
and probabilities (bottom) for the initial state (20), $\theta_i= \pi/2$ and small detuning $\tilde{g_1}=0.2,~\tilde{g_2}=0.202$.}
\label{fig5}
\end{figure}

The bottom parts of Fig. \ref{fig4} and Fig. \ref{fig5} show the probabilities of finding the
qubits in $\vert ee\rangle$, $\vert eg\rangle$, $\vert ge\rangle$ and $\vert
gg\rangle$ states (e.g. $P_{eg}=\vert \langle e g \vert \psi_{QQ}(t)\rangle
\vert^2$). 
We see that the final state belongs to the subspace spanned by $\vert eg\rangle$ and
$\vert ge\rangle$. This is because of the value of
$\theta_i=\pi/2$ and the chosen projection operator. For such $\theta$
the interaction term in (\ref{HQR}) reduces to the form $\tilde{g}\left( a^{\dagger} +
a\right)\sigma_x$  that excites only $\vert e n\rangle$ with $n$ even and
$\vert g m \rangle$ with $m$  odd if we start from $\vert e0\rangle$ and
$\vert g1\rangle$ initial states respectively. Then when the BSM is done the only nonzero
elements, in equation (\ref{pqq}), are $b_1\tilde{a}_0$ and $a_1\tilde{b}_0$.
The relation between the 'occupation probabilities' can be directly translated
into the entanglement of the state: the more one of the probabilities dominates
the other the less entangled is the state and when the probabilities $P_{eg}$
and $P_{ge}$ equal 0.5 the entanglement reaches its maximal value.

The decay rate of the $QR$ system can be estimated as \cite
{blais}
\begin{equation}
\frac{1}{T_{QR}}=\frac{1}{2}\left(\frac{1}{T_Q}+\frac{1}{T_R}\right)
\end{equation}
Assuming the cavity with $Q_f =10^5$ we find $T_R \sim 0.3\mu s$ and
$T_{QR}\sim 0.5 \mu s$. For the cavity with $Q_f = 10^6$ we get $T_R \sim 3 \mu
s$ and $T_{QR} \sim 1.5 \mu s$. The decoherence time of the $QQ$ entangled
state is accordingly $T_{QQ}\sim T_Q\sim1\mu s$. This estimation is in
agreement with the experimental findings \cite{berkeley} that entangled states
do not decohere faster than uncoupled systems.
\\
For the initial state
\begin{equation}
\vert\psi _{QR}(0)\rangle_1 \otimes\vert\psi_{QR}(0)\rangle_2=\vert e
0\rangle_1 \otimes\vert e0\rangle_2 \label{e0e0}
\end{equation}
the situation looks different. The identity of the systems (the same parameters and initial states) 
leads to the striking results. Whenever we perform the BSM we almost always (with some exceptions)
 obtain the maximally entangled qubit-qubit state.
 In order to show some subtleties we take the systems slightly detuned with $\tilde{g_1}=0.2, \tilde{g_2}=0.202$ and
treat $\tilde{g_1}=\tilde{g_2}=0.2$ as a limiting case. Because the two $QR$ systems are almost
identical differing minutely in Rabi frequencies, they evolve to almost the same quantum
 states and even if $QR$'s are
not strongly entangled the BSM gives nearly the same probabilities
$P_{eg}=P_{ge} \sim 0.5$. In consequence, we get almost maximally entangled $QQ$ state
for arbitrary BSM time, except for some moments (in Fig. \ref{fig5} for $\omega_R t \sim 16$ ) at which the norm of the BSM
output approaches zero and the above quantities become undefined. If the BSM
were performed at these moments the entanglement would be unsuccessful. In the
case $\tilde{g_1}=\tilde{g_2}$ the probabilities are always the same and the qubits get
maximally entangled for each BSM time (see B line\ in Fig. \ref{fig7}) with the
exceptions described above. Similar results we have obtained for the initial
state $\vert g1 \rangle_1 \otimes\vert g1 \rangle_2 $.

\section{Decoherence}
Design and construction of quantum devices is always limited by the influence
of environment.  Here, instead of rigorous treatment, developed e.g. for pure
dephasing \cite{lucz,ftc},  we apply the commonly used Markovian approximation
\cite{gardiner} and model the reduced dynamics of the $QR$ system  in terms of
master equation generating complete positive dynamics \cite{alicki}. Following
\cite{blais} we assume that the effect of environment can be included in terms
of two independent Lindblad terms:
\begin{eqnarray}
\dot{\rho}(t)=[L_H-\frac{1}{2}L_{1}-\frac{1}{2}L_{2}]\rho(t)
\end{eqnarray}
where the 'conservative part' is given by
\begin{eqnarray}
L_H(\cdot)=-i[H_Q,\cdot]
\end{eqnarray}
whereas the 'Lindblad dissipators'
\begin{eqnarray}
L_{k}(\cdot)=A^\dagger_kA_k(\cdot)+(\cdot)A^\dagger_kA_k-2A_k(\cdot)A^\dagger_k,
\,\,\,\, k=1,2.
\end{eqnarray}

are expressed in terms of creation and annihilation operators 'weighted' by
suitable lifetimes $A_1=a/\sqrt{T_R}$ and $A_2=\sigma_-/\sqrt{T_{Q}}$. To be
precise we assume  $T_Q\sim 1\mu s$,  $T_R \sim 0.3\mu s$. As the dynamics becomes
non-unitary the system evolves, in general, to the mixed state. The BSM
applied to the density operator of the mixed states  is well defined physical
operation of projection and reduction which can be shown to be completely
positive (see Appendix) and thus applicable to arbitrary ${\rho}$.
\begin{figure}[h]
\centering
\includegraphics [width=0.7\textwidth]{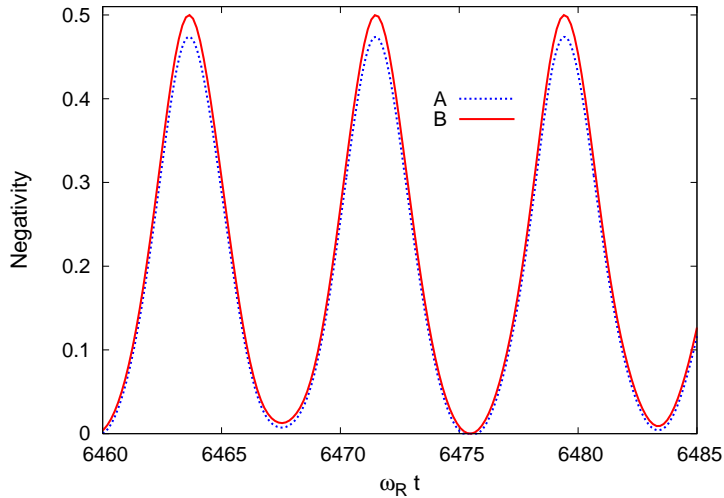}
\caption{(color online) The QQ negativity with (A) and without (B) decoherence for the
initial state (\ref{e0g1}). The parameters are $\theta_i= \pi/2$, $\tilde{g}_i=0.2$,
$T_{R_i}=0.3 \mu s$, $T_{Q_i}=1 \mu s$.} 
\label{fig6}
\end{figure}
In Fig. \ref{fig6} we show the results of the master equation simulations of the negativity
(the line labeled by A in Fig. \ref{fig6}) in comparison with the calculations which neglect decoherence
(the line labeled by B) for the initial state (\ref{e0g1}). The periodicity with the
decoherence included is conserved. For better visibility we present the results
only in a short time period. We see that decoherence decreases slightly the
amplitude of the oscillations. \\
The influence of decoherence on the entanglement of the system starting from
(\ref{e0e0}) (Fig. \ref{fig7}) is much more dramatic. In contrast to the non dissipative
case (B) the result of the BSM depends strongly on the BSM time and the
character of the entanglement becomes quasi-periodic.
\begin{figure}[h]
\centering
\includegraphics [width=0.7\textwidth]{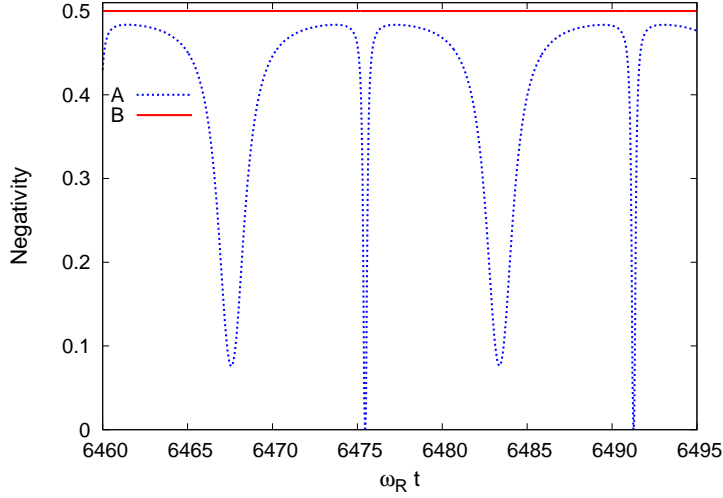}
\caption{(color online) The QQ negativity with (A) and without (B) decoherence for the
initial state (\ref{e0e0}). The parameters are as in Fig. \ref{fig6}.} 
\label{fig7}
\end{figure}
In Fig. \ref{fig8}. we show the decrease of the amplitude of negativity as a function of
time in the larger time scale for both initial conditions. The decrease is
faster for the initial state (\ref{e0g1}) in comparison with that for the initial state (\ref{e0e0}).
\begin{figure}
\centering
\includegraphics [width=0.7\textwidth]{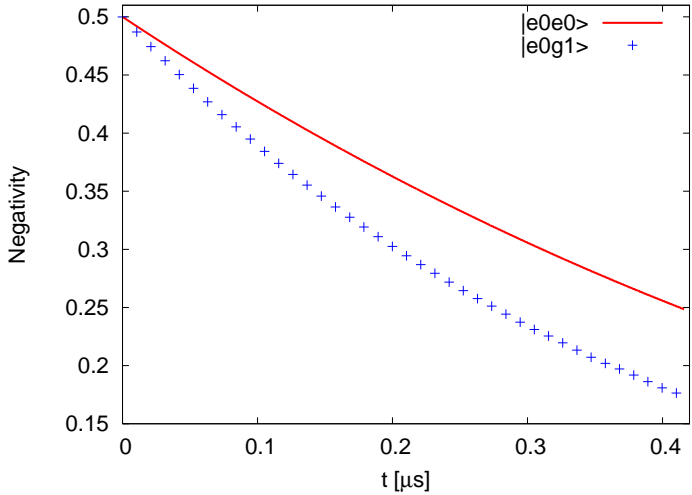}
\caption{(color online) The amplitude of the qubit-qubit negativity plotted as a function of BSM
time for two different initial states. The parameters are as in Fig. \ref{fig6}.}
\label{fig8}
\end{figure}

\section{Conclusions}
We investigated a mechanism for creation of entanglement of two qubits, each
interacting with a single mode electromagnetic field coming from independent
sources. This interaction leads to two independent entangled qubit-field states
and that BSM performed on the electromagnetic field modes projects the qubits
onto an entangled state. Thus we discussed transfer of quantum information
between systems having different physical nature and defined in Hilbert spaces
of different dimensions.

   In the first part of the  paper we have dealt  with the pure states which is
justified to some extent by estimated relatively long decoherence times. The
discussed systems offer the advantage of reaching a strong coupling regime
between light and matter. We have checked that the Jaynes-Cummings model,
valid for weaker $QR$ coupling \cite{geller}, gives the results in agreement
with our calculations for $g \le 0.03 \omega_R$. Assuming reasonable values of
parameters we found that the strong coupling regime ($T_\Omega ^{-1}\gg
T_R^{-1},T_Q^{-1}$) can be realized and coherent manipulations of qubits
(especially with the quantum error correction technique) and maximally entangled
qubit-qubit states are possible. \\
Analyzing the dynamics of the system in the presence of decoherence we found
that the observation of coherent phenomena and in particular the generation of
highly entangled states is still possible.

It seems that entanglement of distant qubits by swapping can have some
advantages over standard schemes of setting up entanglement that rely on
generating entangled subsystems at a point and supplying them to distant areas.
The qubits emerge entangled despite the fact that they never interacted in the
past and therefore they do not influence each other by disturbing the single
qubit features. They can be at much larger distances as the scheme does not
depend essentially on the distance between them. The degree of entanglement
depends on the moment at which the BSM was performed. Verifying experimentally
that two qubits are unambiguously entangled is a difficult task requiring
sophisticated methods such as e.g. quantum state tomography \cite{steffen}. The
solid state qubits and their entanglement discussed in this paper can be scaled
to a larger set of quantum bits \cite{bose}. It can be of interest in the study
of fundamental laws of quantum mechanics and can be useful in quantum
information processing and quantum communication. It seems that the
experimental realization of the presented model considerations may be performed
with currently available technologies. Following \cite{hor4}, we hope that
'what is predicted by quantum formalism must occur in the laboratory'. Sooner
or later.
\ack
J.D. thanks Marcin Mierzejewski for stimulating discussions concerning
numerical methods applied in this paper. Work supported by the Polish Ministry
of Science and Higher Education under the grant N 202 131 32/3786 and by RITA
-CT-2003-506095.

\section*{Appendix}
We will prove that the transformation described by (\ref{rhoqq}) and
(\ref{bell}) which was defined for pure states, makes sense also for an
arbitrary mixed state of the two-qubit-field ($QRQR$) system, ie.\ it is
described by a completely positive operator transforming an arbitrary density
matrix of the full system into a density matrix of two-qubit ($QQ$) system.
Although it is easy to understand on a purely physical basis (transformation
consists of a measurement and a reduction to a subsystem), it is instructive to
give an explicit proof of the statement. As a bonus we will easily find an
explicit Kraus form of the transformation in question.

Let
\begin{equation}\label{rho}
\rho=\sum_{\substack{\mu k\nu l\\ \varsigma m \tau n}}
\rho^{\mu k\nu l}_{\varsigma m \tau n}\kb{\mu k\nu l}{\varsigma m \tau n},
\end{equation}
where, cf.\ (\ref{rhopsi}),
\begin{equation}\label{pure}
\ket{\mu k\nu l}=\ket{\mu}_1\otimes\ket{k}_1\otimes\ket{\nu}_2\otimes\ket{l}_2
=\ket{\psi_{QR}}_1\otimes\ket{\psi_{QR}}_2
\end{equation}
for $\mu,\nu\in\{g,e\}$, $k,l\in\{0,1,\ldots\}$ form a basis of pure states for
the full system.

For the moment let us consider only the Jaynes-Cummings approximation where we
take into account only the modes $\ket{0}$ and $\ket{1}$ of the electromagnetic
field, hence all Latin indices in (\ref{rho}) and (\ref{pure}) take the values
$0,1$ only. In this case density matrices of the $QRQR$ system act in the
$16$-dimensional complex space, $\mathcal{H}_1=\mathbb{C}^{16}$, and as such
form a subset of the $16\times 16$-dimensional complex linear space.
Analogously, density matrices of the $QQ$ system, acting in the $4$-dimensional
complex space, $\mathcal{H}_2=\mathbb{C}^{4}$, form a subset of the $4\times
4$-complex space. The transformation (denoted in the following by $\Lambda$)
described by (\ref{rhoqq}) regarded on the whole $256$-dimensional complex
space transforms it into the $16$-dimensional one. Straightforward calculations
give
\begin{eqnarray}\label{coord}
\Lambda(\rho)&=:&\sigma=
\sum_{\substack{\mu,\nu\in\{g,e\} \\ \varsigma,\tau\in\{g,e\}}}
\sigma^{\mu\nu}_{\varsigma\tau}\kb{\mu\nu}{\varsigma\tau}, \\
\sigma^{\mu\nu}_{\varsigma\tau}&=&\frac{1}{2}\left(
\rho^{\mu 0 \nu 1}_{\varsigma 0 \tau 1}-
\rho^{\mu 1 \nu 0}_{\varsigma 0 \tau 1}-
\rho^{\mu 0 \nu 1}_{\varsigma 1 \tau 0}+
\rho^{\mu 1 \nu 0}_{\varsigma 1 \tau 0}\right),
\end{eqnarray}
where $\ket{\mu\nu}:=\ket{\mu}_1\otimes\ket{\nu}_2$ form a basis of pure states
of the $QQ$ system. In the following we will need only
\begin{eqnarray}\label{Lambda2}
&&\Lambda\big(\kb{\mu k\nu l}{\varsigma m \tau n}\big)=
\nonumber \\
&=&
\frac{1}{2}\big(
\delta_{0k}\delta_{1l}\delta_{0m}\delta_{1n}-
\delta_{1k}\delta_{0l}\delta_{0m}\delta_{1n}
\nonumber \\
&-&
\delta_{0k}\delta_{1l}\delta_{1m}\delta_{0n}+
\delta_{1k}\delta_{0l}\delta_{1m}\delta_{0n}
\big)\kb{\mu\nu}{\varsigma\tau}
\end{eqnarray}
To check the complete positivity of $\Lambda$ we use the Choi-Jamiolkowski
isomorphism defined as \cite{jam}
\begin{equation}\label{Jam1}
\mathcal{J}(\Lambda)=(\Lambda\otimes\mathbb{I}_1)(P_+),
\end{equation}
where $\mathbb{I}_1$ is the identity operator on the $256$-dimensional space
and $P_+$ is a maximally entangled state on the
$\mathcal{H}_1\otimes\mathcal{H}_1$ space
\begin{equation}\label{maxent}
P_+=\kb{\Phi_+}{\Phi_+}, \quad \ket{\Phi_+}=\sum\limits_{\mu k\nu l}
\ket{\mu k\nu l}\otimes\ket{\mu k\nu l}.
\end{equation}
According to the Choi theorem \cite{choi}, $\Lambda$ is completely positive if
and only if $\mathcal{J}(\Lambda)$ is a positive-definite operator. Applying
(\ref{Jam1}) and (\ref{Lambda2}) to (\ref{maxent}) we get
\begin{equation}\label{J2}
\mathcal{J}(\Lambda)=\kb{\Phi}{\Phi},
\end{equation}
where
\begin{equation}\label{Phi}
\ket{\Phi}=\frac{1}{\sqrt{2}}\sum\limits_{\mu\nu}
\ket{\mu\nu}\otimes\big(\ket{\mu 0\nu 1}-\ket{\mu 1\nu 0}\big).
\end{equation}article
Hence $\mathcal{J}(\Lambda)$ is a projection and as such a positive definite
operator, consequently $\Lambda$ is completely positive.

The obtained results allow to write explicitly the so called Kraus form of
$\Lambda$,
\begin{equation}\label{kraus0}
\Lambda(\rho)=\sum_n A_n\rho A_n^\dagger,
\end{equation}
where $A_n$ are $\mathrm{dim}\mathcal{H}_2\times \mathrm{dim}\mathcal{H}_1=4\times 16$ matrices.
To this end \cite{arrighi} we have to perform the spectral decomposition of the positive
definite operator $\mathcal{J}(\Lambda)$
\begin{equation}\label{kraus2}
\mathcal{J}(\Lambda)=
\sum\limits_\mu d_n\kb{\chi^\prime_n}{\chi^\prime_n}.
\end{equation}
Since $d_n$ are positive, we can rescale the eigenvectors
\begin{equation}\label{kraus3}
\ket{\chi_n}:=\sqrt{d_n}\ket{\chi^\prime_n}.
\end{equation}
Now the operators $A_n$ can be found in the form
\begin{equation}\label{kraus5}
A_n:=\big(\mathbb{I}_2\otimes\bra{\Phi_+}\big)
\big(\ket{\chi_n}\otimes\mathbb{I}_1\big),
\end{equation}
where $\mathbb{I}_2$ is the identity on $\mathcal{H}_2$. The above formula
should be properly understood. Observe that since $\ket{\chi_n}$ is an element
of $\mathcal{H}_2\otimes\mathcal{H}_1$, it has the form $\ket{\chi_n}=
\sum_i\ket{\phi_{n,i}}\otimes\ket{\xi_{n,i}}$, where
$\ket{\phi_{n,i}}\in\mathcal{H}_2$, $\ket{\xi_{n,i}}\in\mathcal{H}_1$, whereas
$\bra{\Phi_+}=\sum\bra{\mu k\nu l}\otimes\bra{\mu k\nu l}$. Hence
\begin{eqnarray}\label{kraus5a}
A_n&=&\big(\mathbb{I}_2\otimes\bra{\Phi_+}\big)
\left(\ket{\chi_n}\otimes\mathbb{I}_1\right)
\nonumber \\
&=&\big(\mathbb{I}_2
\otimes\sum_{\mu k\nu l}\bra{{\mu k\nu l}}\otimes\bra{{\mu k\nu l}}\big)
\big(\sum_i\ket{\phi_{n,i}}\otimes\ket{\xi_{n,i}}\otimes\mathbb{I}_1\big)
\nonumber \\
&=&\sum_{i,{\mu k\nu l}}\bk{{\mu k\nu l}}{\xi_{n,i}}\,\kb{\phi_{n,i}}{{\mu k\nu l}}.
\end{eqnarray}
In our case $\mathcal{J}(\Lambda)$ has only one non vanishing eigenvalue
corresponding to the eigenvector
$\ket{\chi_1}=\ket{\Phi}=\sum_{\mu\nu}\ket{\phi_{1,\mu\nu}}\otimes\ket{\xi_{1,\mu\nu}}$.
Hence $\ket{\phi_{1,\mu\nu}}=\ket{\mu\nu}$ and $\ket{\xi_{1,\mu\nu}}=\big(\ket{\mu
0\nu 1}-\ket{\mu 1\nu 0}\big)/\sqrt{2}$. Using (\ref{Phi}) and (\ref{kraus5a}) we
obtain finally:
\begin{equation}\label{A}
A=\frac{1}{\sqrt{2}}\sum_{\mu\nu}\big(\kb{\mu\nu}{\mu 0\nu 1}-\kb{\mu\nu}{\mu 1\nu 0}\big)
\end{equation}
A short calculation shows that indeed, cf.\ (\ref{coord}),
\begin{equation}\label{ch1}
\Lambda(\rho)=A\rho A^\dagger.
\end{equation}

The calculations do not change considerably if we go beyond the Jaynes-Cummings
approximation, by taking into account arbitrary finite numbers of photons in
each cavity. In fact, in this case, the only difference consists of extending
all summations over the number of photons from two terms corresponding to 0 and
1 to the desired numbers of cavity excitations which we would like to regard.
The final results (\ref{A}) and (\ref{ch1}) remain unaltered. The situation is
more subtle if we want to take into account the infinite number of possible
photonic excitations of the cavity modes. The corresponding cavity Hilbert
space becomes now infinite-dimensional and a straightforward generalization of
the Choi-Jamio{\l}kowski isomorphism does not exist -- one has to resort to
slightly more involved procedures to investigate directly the complete
positivity \cite{grabowski}. It is, however, not really needed in our case.
As it is easy to check, the final result (\ref{A}), (\ref{ch1}) is
correct also in the infinite-dimensional setting.

\end{document}